\documentclass[intlimits,twoside,a4paper]{article}

\usepackage{amsmath,amssymb}
\usepackage{graphicx}
\usepackage{wrapfig}

\usepackage[T2A]{fontenc}
\usepackage[cp1251]{inputenc}

\usepackage{cmpj2}

\issue{2014}{17}{2}{23704}
\doinumber{10.5488/CMP.17.23704}

\title[The effect of interface phonons on operating
electron states\ldots]%
{The effect of interface phonons on operating electron states in
three-barrier resonant tunneling structure as an active region
of quantum cascade detector}
\author[M.V.~Tkach \textsl{et al.}]{M.V.~Tkach\footnote{E-mail: ktf@chnu.edu.ua}\, ,
Ju.O.~Seti, Y.B.~Grynyshyn, O.M.~Voitsekhivska}
\address{Chernivtsi National University, 2 Kotsubynsky St.,
58012 Chernivtsi, Ukraine}

\date{Received April 16, 2014, in final form May 17, 2014}
\authorcopyright{M.V.~Tkach, Ju.O.~Seti, Y.B.~Grynyshyn, O.M.~Voitsekhivska, 2014}

\begin{document}

\maketitle

\begin{abstract}
The Hamiltonian of electrons interacting with interface phonons in
three-barrier resonant tunneling structure is established using
the first principles within the models of effective mass and
polarization continuum. Using the Green's functions method, the
temperature shifts and decay rates of operating electron states
are calculated depending on geometric design  of three-barrier
nano-structure GaAs/Al$_{x}$Ga$_{1-x}$As which is an active region
of quantum cascade detector. It is established that independently
of the temperature, the energy of quantum transition during the
process of electromagnetic field absorption is a nonlinear
weakly varying function of the position of the inner barrier
with respect to the outer barriers of the structure.

\pacs 78.67.De, 63.20.Kr, 72.10.Di
\end{abstract}

\section{Introduction}

Quantum cascade detectors (QCD) have been investigated for over a
decade but a growing attention to these devices is still observed \cite{1,2,3,4}. From
practical point of view, such an interest is caused by the unique
characteristics of QCD. Occupying the whole infrared and terahertz
frequency range of electromagnetic waves, these devices can
operate in a wide range of temperatures (from cryogenic to room
ones) and so on. From theoretical point of view, the interest
is caused by the fact that the basis of QCD functional elements
are the open quasi-two-dimensional resonant tunneling structures
(RTS). The physical properties of electronic transport in these
structures are not still clear enough.

The theory of spectral parameters and dynamic conductivity of
electrons in open RTS is well developed \cite{5,6,7} without taking
into account the electron-phonon interaction. The obtained results well
correlate with the experimental data \cite{3}. However, from
physical considerations it is clear that the effect of phonons
can be neglected only at cryogenic temperatures when the
average occupation numbers of phonon states are small and
electron-phonon binding is weak. Studying the nano-structures with
strong binding or at high temperatures (modern QCDs operate
at room temperatures \cite{3}), one should consider the electron-phonon
interaction.

The theory of electron-phonon interaction in spherical,
cylindrical and plane closed single \cite{8,9} and multi-shell
nano-heterosystems \cite{10,11,12,13,14,15} has been developed for a long
time using the models of effective mass and dielectric continuum.
It was established that, contrary to the massive
three-dimensional systems, the so-called interface phonons
(I-phonons) exist in low-dimensional nano-heterosystems (besides
the confined polarization phonons). The effect of I-phonons
increases when the thickness of nano-layers decreases.

The effect of phonons on the transport properties of electronic current
through RTS was mainly investigated for the two- and three-barrier
nano-structures \cite{5,6,7}. Studying the probabilities of quantum
transitions using the Fermi golden rule, it is enough to use the
Hamiltonian of electron-phonon interaction in the representation
of the second quantization over the phonon variables only, obtained by
Mori and Ando \cite{8} for a double heterostructure.

In this study, we investigate the electron spectrum renormalized
due to I-phonons in three-barrier RTS which is an active region of
QCD. The Hamiltonian of electron-I-phonon system is obtained in
the representation of occupation numbers over all variables. It is
further used in the method of temperature Green's functions in
order to study the shifts and decay rates of two lower electron
states - the operating states of QCD active region. This makes it possible
to study in detail the effect of various
mechanisms of electron-I-phonon interaction on the parameters of
two operating electron states depending on the design of
three-barrier RTS at cryogenic and room temperatures.

\section{Hamiltonian and Fourier image of Green's
function of the system of electrons interacting with interface
polarization phonons in a three-barrier nano-structure}

\begin{wrapfigure}{i}{0.5\textwidth}
\centerline{
\includegraphics[width=0.45\textwidth]{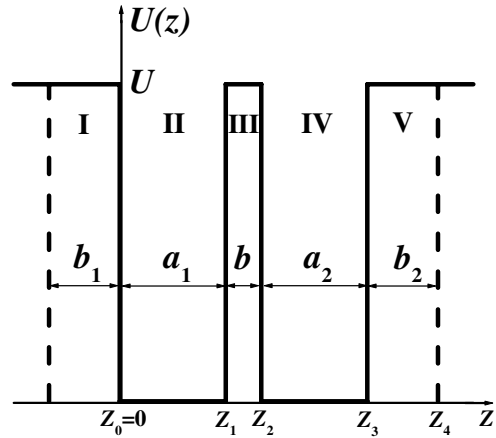}
}
\caption{ Potential energy profile of closed three-barrier RTS
(solid line). The boundaries of outer barriers with the widths
($b_{1}$; $b_{2}$) for the corresponding open system
(dashed lines).} \label{fig1}
\end{wrapfigure}

The theory of spectral parameters (resonance energies and decay
rates) and dynamic conductivity of electrons in three-barrier RTS
without taking into account the electron-phonon interaction was
developed in detail in \cite{5}. It was established that when the
widths of nano-structure outer barriers were bigger than 3--4~nm,
the resonance energies in open and closed models were almost the same
and the resonance widths were two-three orders smaller than the
energies. Considering the widths of outer barriers of
a three-barrier RTS (the active bands of experimental QCD) as rather
big (3--6~nm) \cite{6}, we develop the theory of electron-I-phonon
interaction using the model of closed three-barrier RTS (figure~\ref{fig1})
with fixed effective masses $m\left(z\right)=\left\{m_\textrm{w} \, ({\rm
II,}\, {\rm \; IV});\, \, \, \, \, m_\textrm{b} \, ({\rm I,}\, \, {\rm
III,}\, \, {\rm V})\right\}$ and rectangular potential energy
profile neglecting the small decay rate
\begin{equation} \label{EQ:1}
U\left(z\right)=\left\{
\begin{array}{llll}
{0}, &   0\leqslant z\leqslant a_{1} \ ({\rm II}), & a_{1} +b\leqslant z\leqslant a_{1} +b+a_{2} \ ({\rm IV}), &
\\ {U}, & -\infty \leqslant z\leqslant 0 \  ({\rm I}), &
a_{1} \leqslant z\leqslant a_{1} +b \ ({\rm III}), & a_{1} +b+a_{2} \leqslant
z\leqslant \infty \ ({\rm V}).
\end{array}\right.
\end{equation}

Expressing the electron wave function in the form
\begin{equation} \label{EQ:2}
\Psi _{E\vec{k}} (\vec{r})=\frac{1}{\sqrt{S} }
\re^{\ri\vec{k}\vec{\rho }} \Psi _{E} (z)\, ,
\end{equation}
where $\vec{k}$ and $\vec{\rho }$ are quasi-momentum and
radius-vector of electron in the plane $xOy$ and ${S}$
is the square of the main region in this plane. For ${z}$-th
component of this function, we obtain the Schrodinger equation
\begin{equation} \label{EQ:3}
\left\{-\frac{\hbar ^{2} }{2} \frac{\rd}{\rd z}
\frac{1}{m\left(z\right)} \frac{\rd}{\rd z}
+U\left(z\right)\right\}\Psi _{E} \left(z\right)=E\Psi _{E}
\left(z\right) .
\end{equation}

The complete electron energy in the region under the barrier ($E
\leqslant U$) consists of two terms
\begin{equation} \label{EQ:4}
E_{n\vec{k}} =E_{n} +\frac{\hbar ^{2} {k}^{2} }{2m_{n}^{*} } \,.
\end{equation}

Here, $\hbar ^{2} k^{2} /(2m_{n}^{*} )$ is the kinetic energy of
electron moving in the plane perpendicular to ${Oz}$ axis. It
is determined, as in \cite{16}, by the effective mass correlated over
the RTS
\begin{equation} \label{EQ:5}
\frac{1}{m_{n}^{*} } =\int _{-\infty }^{\infty }\frac{\left|\Psi
_{n} \left(z\right)\, \right|^{2} \rd z}{m\left(z\right)}  \, ,
\end{equation}
where $\Psi _{n} \left(z\right)$ wave functions  are the solutions
of one-dimensional
stationary equation (\ref{EQ:3})
\begin{equation} \label{EQ:6}
\Psi _{n} \left(z\right)=\left\{
\begin{array}{l} {\ \sum\limits_{j=2,4}\Psi _{jn} \left(z\right)
=\sum\limits_{j=2,4}\left(A_{jn} \cos k_{n} z+B_{jn} \sin k_{n} z\right)\, ,} \\[1ex]
{\sum\limits_{j=1,3,5}\Psi _{jn} \left(z\right)=\sum\limits_{j=1,3,5}
\left(A_{jn} \re^{\chi _{n} z} +B_{jn} \re^{-\chi_{n} z} \right)\,  .}
\end{array}\right.
\end{equation}
Here,
\begin{equation} \label{EQ:7}
k_{n} =\hbar ^{-1} \sqrt{2m_\textrm{w} E_{n} } \, ,
\qquad
 \chi _{n} =\hbar ^{-1} \sqrt{2m_\textrm{b} \left|U-E_{n}
\right|} =\sqrt{2m_\textrm{b} U\hbar ^{-2} -k_{n}^{2} m_\textrm{b} /m_\textrm{w} }\, .
\end{equation}

The discrete energy spectrum $E_{n} $ and coefficients $A_{jn}$,
 $B_{jn} $ are fixed by fitting conditions
\begin{equation} \label{EQ:8}
\left\{\begin{array}{l} {\Psi _{jn} (z)\big|_{z=z_{j} }
=\Psi _{j+1,\, n} (z)\big|_{z=z_{j} }  }\, , \\[2ex]
{\displaystyle \frac{1}{m_{j} }  \frac{\partial \Psi _{jn}
}{\partial z} \bigg|_{z=z_{j} } =\frac{1}{m_{j+1} }
\frac{\partial \Psi _{j+1,\, n} }{\partial z} \bigg|_{z=z_{j} }\, ,
\qquad {j}={1,2,3,4}}
\end{array}\right.
\end{equation}
together with the condition that the wave function vanishes at $z\to
\pm \infty$  \ ($B_{1n} =A_{5n} =0$) and the normality condition
\begin{equation} \label{EQ:9}
\int _{-\infty }^{\infty }\Psi _{n}^{*} (z)\Psi _{n'} (z)\rd z
=\delta _{nn'} \,.
\end{equation}

In the region above the barrier ($E\geqslant U$), the energy of electron
longitudinal movement is continuous. Thus, introducing the
longitudinal quasi-momentum $k_{z} $, it is written as $E_{k_{z} }
=\hbar ^{2} k_{z}^{2} /(2m_\textrm{w} )$. Finally, the complete energy has
the form:
\begin{equation} \label{EQ:10}
E_{k_{z} \vec{k}} =E_{k_{z} } +\frac{\hbar ^{2} k^{2} }{2m_\textrm{w}} \,.
\end{equation}

Now, the solution of equation (\ref{EQ:3}) for the wave function
$\Psi_{k_{z}} (z)$ becomes the expression (\ref{EQ:6}) with $k_{n} \to
k_{z}$ , $\chi _{n} \to \ri \chi$ . The fitting conditions
\eqref{EQ:8} are valid at $n\to k_{z} $, and the normality
condition, similarly to the de Broglie wave, is written as follows:
\begin{equation} \label{EQ:11}
\int_{-L/2}^{L/2}\left|\Psi _{k_{z}} (z)\right|^{2} \rd z=1,
\end{equation}
where $\Psi _{k_{z}} (z)$ function satisfies the periodic
condition $\Psi _{k_{z}} (-L/2)=\Psi _{k_{z}} (L/2)$
at a big span of the main region having the length $L$. All
coefficients ($A_{jk_{z} }$, $B_{jk_{z} } $) are defined from
these conditions and the wave function $\Psi _{k_{z}} (z)$ is
obtained.

Introducing the generalized index
$\tilde{n}=\left\{\begin{smallmatrix} n, \ & E \leqslant U \\ k_{z},& E\geqslant U \end{smallmatrix}\right\}$,
for the compact analytics, we perform a transition to the representation
of the second quantization using the quantized wave function
\begin{equation} \label{EQ:12}
\stackrel{\frown}{\Psi }(\vec{r})=\sum _{\tilde{n}\vec{k}}\Psi
_{\tilde{n}\vec{k}} (\vec{\rho
},z)\stackrel{\frown}{a}_{\tilde{n}\vec{k}}  =\sum
_{\vec{k}}\left[\sum _{n}\Psi _{n}  (\vec{\rho },z)a_{n\vec{k}}
+\sum _{k_{z} }\Psi _{k_{z} }  (\vec{\rho },z)a_{k_{z} \vec{k}}
\right]
\end{equation}
and obtain the Hamiltonian of uncoupling electrons in the
representation of their occupation numbers
\begin{equation} \label{EQ:13}
\stackrel{\frown}{H}_\textrm{e} =\sum
_{\tilde{n},\vec{k}}E_{\tilde{n}\vec{k}}\, a_{\tilde{n}\vec{k}}^{+}
a_{\tilde{n}\vec{k}}
\end{equation}
with the electron spectrum {$E_{\tilde{n}\vec{k}} $},
creation ($a_{\tilde{n}\vec{k}}^{+} $) and annihilation
($a_{\tilde{n}\vec{k}} $) Fermi operators of electron states,
satisfying the anti-commutative relationships.

It is well known \cite{8,9,10} that in the dielectric continuum
model, the phonon spectra and potential of polarization field
{$\Phi (\vec{r})$} are obtained from the following equation
\begin{equation} \label{EQ:14}
\varepsilon _{j} (\omega )\nabla ^{2} \Phi (\vec{r})=0\, ,
\end{equation}
where $\varepsilon _{j} (\omega )$ is dielectric constant of
${j}$-th layer of a nano-structure composed of two materials
\begin{equation} \label{EQ:15}
\varepsilon _{j} (\omega )=\varepsilon _{j\infty } \frac{\omega
^{2} -\omega _{Lj}^{2} }{\omega ^{2} -\omega _{Tj}^{2} }  \, .
\end{equation}
Here, $\varepsilon _{j\infty } $ is high-frequency dielectric
constant, $\omega _{Lj}$,  $\omega _{Tj}$ are the frequencies
of longitudinal (${L}$) and transversal (${T}$) phonons
of the bulk material creating  ${j}$-th layer of
nano-structure.

The spectrum and potential of polarization field of interface
phonons is obtained, according to \cite{9}, from the equation (\ref{EQ:14}) if
\begin{equation} \label{EQ:16}
\nabla ^{2} \Phi (\vec{r})=0.
\end{equation}
Thus, the solution of this equation is the potential
\begin{equation} \label{EQ:17}
\Phi (\vec{r})=\sum _{j,\vec{q}}C(q) {\varphi}_{j}
(q,z)\re^{\ri\vec{q}\vec{\rho }} \, ,
\end{equation}
where $\vec{q}$, $\vec{\rho }$ are two-dimensional vectors and
functions
\begin{equation} \label{EQ:18}
{\varphi }_{j} (q,z)=\alpha _{j} \re^{-qz} +\beta _{j} \re^{qz} \, ,\qquad j=1,\ldots,5
\end{equation}
satisfy the system of equations obtained from the fitting
conditions for the intensity and induction of polarization field
\begin{equation} \label{EQ:19}
\left\{
\begin{array}{l}
{\varphi _{j} (q,z_{j} )=\varphi _{j+1}
(q,z_{j} )}, \\[1ex]
\displaystyle{\varepsilon _{j} (\omega
)\frac{\partial \varphi _{j} (q,z)}{\partial z} \bigg|_{z=z_{j} }
= \varepsilon _{j+1} (\omega )\frac{\partial \varphi _{j+1}
(q,z)}{\partial z} \bigg|_{z=z_{j} } }, \end{array}\right. \qquad  j=1,2,3,4
\end{equation}
and considering that at $z\to \pm \infty $$ $
\begin{equation} \label{EQ:20}
 \varphi _{1} (q,z)\big|_{z\to -\infty } =\varphi
_{5} (q,z)\big|_{z\to \infty } =0.
\end{equation}

Within the transfer-matrix method \cite{11}, the coefficients $\alpha
$${}_{j}$, $\beta $${}_{j}$ and the potential of
polarization field $\Phi (\vec{r})$ are obtained from the system
of equations (\ref{EQ:19}), (\ref{EQ:20}). The condition of nontrivial solution
determines the dispersion equation
\begin{equation} \label{EQ:21}
\prod _{j=1}^{5} \left(
\begin{array}{ll}
\left[1+\frac{\varepsilon _{1} (\Omega )}{\varepsilon _{0}
(\Omega )} \right] &  \left[1-\frac{\varepsilon _{1} (\Omega
)}{\varepsilon _{0} (\Omega )} \right]\re^{-2qz_{j-1} }  \\[1ex]
\left[1-\frac{\varepsilon _{1} (\Omega )}{\varepsilon _{0} (\Omega
)} \right]\re^{2qz_{j-1} } & \left[1+\frac{\varepsilon _{1}
(\Omega )}{\varepsilon _{0} (\Omega )} \right]
\end{array}
\right)
=\left(\begin{array}{ll} 1 & 0 \\ 0 & 1 \end{array}\right).
\end{equation}
Here, $\varepsilon _{0} (\Omega )$ and $\varepsilon _{1} (\Omega )$
are the dielectric constants in the wells and barriers,
respectively.

 Its solutions $\Omega _{\lambda \vec{q}} =\hbar \omega
_{\lambda \vec{q}} $ define the energy spectrum of interface
phonons. For the nondegenerated case, the number of phonon modes
($\lambda $) is equal to the twice number of all interfaces
between nano-structure layers.

Quantizing the polarization field using the known quantum mechanics method \cite{9}, we obtain the Hamiltonian of interface phonons
\begin{equation} \label{EQ:22}
\stackrel{\frown}{H}_\textrm{I} =\sum _{\lambda,\vec{q}}\Omega _{\lambda
\vec{q}}  \left(b_{\lambda \vec{q}}^{+} b_{\lambda \vec{q}}
+\frac{1}{2}
\right)\, ,\qquad \lambda =1,\ldots,8
\end{equation}
and the Hamiltonian of electron-I-phonon interaction
\begin{equation} \label{EQ:23}
\stackrel{\frown}{H}_\textrm{e-I} =-e\Phi (\vec{\rho },z)=-\sum _{\lambda
,\vec{q},j}eC_{\lambda } (q) \, \varphi _{j}
(\lambda,q,z)\re^{\ri\vec{q}\vec{\rho }} \left(b_{\lambda \vec{q}}
+b_{\lambda ,-\vec{q}}^{+} \right)
\end{equation}
in coordinate representation over the electron variables
$(\vec{\rho },z)$ and in the representation of occupation numbers
over the phonon variables with operators $b_{\lambda \vec{q}}^{+}$,  $b_{\lambda \vec{q}} $, satisfying commutative
relationships and with the known coefficients $C_{\lambda } (q)$ and
functions $\varphi _{j} (\lambda,q,z)$.

Performing the transition to the representation of electron
occupation numbers in (\ref{EQ:23}) using the quantized wave function (\ref{EQ:12}),
we obtain the Hamiltonian of electron-I-phonon interaction in the
representation of the second quantization over all variables of the
system
\begin{equation} \label{EQ:24}
\stackrel{\frown}{H}_\textrm{e-I} =\sum _{\small{\begin{smallmatrix}
{\tilde{n}_{1}, \tilde{n},\vec{k}}\\ {\lambda ,\vec{q}}
\end{smallmatrix}}}F_{\tilde{n}_{1}\tilde{n}} (\lambda
,\vec{q})a_{\tilde{n}_{1} \vec{k}+\vec{q}}^{+}
a_{\tilde{n}\vec{k}}^{} \left(b_{\lambda \vec{q}} +b_{\lambda
,-\vec{q}}^{+} \right),
\end{equation}
where binding functions
\begin{equation} \label{EQ:25}
F_{\tilde{n}_{1} \tilde{n}} (\lambda ,\vec{q})=-\sqrt{\frac{8\pi
\Omega a_\textrm{b} \textrm{Ry} }{L^{2} qN(\lambda ,q)} } \sum _{j=1}^{5}\, \int
_{z_{j-1} }^{z_{j} }\Psi _{\tilde{n}_{1} j}^{*} (z)\Psi
_{\tilde{n}j} (z) \left[\alpha _{j} (\Omega _{\lambda q} )\re^{-qz}
+\beta _{j} (\Omega _{\lambda q} )\re^{qz} \right] \, \rd z
\end{equation}
contain the normality coefficient
\begin{equation} \label{EQ:26}
N(\lambda ,q)=\sum _{j=1}^{5}\omega _{_{\lambda =1} }^{(q=0)}
 \frac{\partial \varepsilon _{j(\omega )} }{\partial \omega
} \bigg|_{\omega =\omega _{\lambda } (q)} \left[\beta _{j}^{2}
(\Omega _{\lambda \vec{q}} )\left(\re^{2qz_{j} } -\re^{2qz_{j-1} } \right)
-\alpha_{j}^{2} \left(\Omega _{\lambda \vec{q}} \right)\left(\re^{-2qz_{j} } -\re^{-2qz_{j-1}
} \right)\right] .
\end{equation}
Here, $\textrm{Ry}={e^{2} / 2a_\textrm{b} } $, $\Omega =\hbar
\omega _{_{\lambda =1} }^{(q=0)} $, $a_\textrm{b} $ is Bohr radius.
Integral in (\ref{EQ:25}) is analytically calculated but we do not present
it due to its sophisticated form.

We should note that at $q \rightarrow 0$, the binding function $F_{\tilde{n}_{1} \tilde{n}} (\lambda, \vec{q})\sim q^{-1/2}$ and, thus, a further integration in MO is not
 divergent.

 The obtained Hamiltonian of electron-I-phonon system in
three-barrier RTS
\begin{equation} \label{EQ:27}
H=H_\textrm{e} +H_\textrm{I} +H_\textrm{e-I}
\end{equation}
allows us to calculate the Fourier-image of electron Green's
function in the quasi-stationary part of the spectrum according to
the rules of Feynman-Pines diagram technique \cite{9}, at the finite
temperature, when Dyson equation is valid
\begin{equation} \label{EQ:28}
G_{n} (\vec{k},\hbar \omega )=\left[ \hbar \omega -E_{n\vec{k}}
-M_{n} (\hbar \omega ,\vec{k})\right]^{-1}
\end{equation}
with mass operator (MO) $M_{n} (\hbar \omega ,\vec{k})$ calculated
(due to the weak electron-I-phonon binding) in one-pho\-non
approximation ($\eta \to \pm 0$)
\begin{equation} \label{EQ:29}
M_{n} (\hbar \omega ,\vec{k})=\sum _{\tilde{n}_{1}, \lambda,
\vec{q}}F_{{n}\tilde{n}_{1} }^{*}  (\lambda
,\vec{q})F_{\tilde{n}_{1}{n}} (\lambda ,\vec{q})\left[\frac{1+\nu
_{\lambda \vec{q}} }{\hbar \omega -E_{\tilde{n}_{1} }
(\vec{k}+\vec{q})-\Omega _{\lambda \vec{q}} +\ri\eta } +\frac{\nu
_{\lambda \vec{q}} }{\hbar \omega -E_{\tilde{n}_{1} }
(\vec{k}+\vec{q})+\Omega _{\lambda \vec{q}} +\ri\eta } \right] ,
\end{equation}
where $\nu _{\lambda \vec{q}} =\left[\re^{\Omega _{\lambda \vec{q}}
/kT} -1\right]^{-1} $ is the average number of I-phonons
occupation numbers.

Further, using this MO, we study the contributions of different mechanisms of electron-I-phonon
interaction into renormalized spectral parameters (energy shifts ($\Delta $${}_{n}$) and decay
rates~($\Gamma _{n} $)) of n-th electron state. In experiments \cite{1,2,3}, the electrons move perpendicularly
to the planes of three-barrier RTS, thus in MO (\ref{EQ:25}) we put $\vec{k}=0$ and neglect the frequency
dependence of MO in the vicinity of $E_{n} $ energies taking into account a weak electron phonon binding
(further proven by numeric calculations). To distinguish the role of different mechanisms of electron-I-phonon
interaction, one should extract the real and imaginary part in MO
\begin{equation} \label{EQ:30}
M_{n} (\vec{k}=0,\, \hbar \omega =E_{n} )=\textrm{Re}M_{n} (\vec{k}=0,\hbar
\omega =E_{n} )+\ri \textrm{Im}M_{n} (\vec{k}=0,\hbar \omega =E_{n} )=\Delta
_{n} -\ri\Gamma _{n} /2 \, ,
\end{equation}
as well as the terms describing the partial contributions of
I-phonons due to interaction with electrons from different states
\begin{equation} \label{EQ:31}
\Delta _{n} =\Delta _{nn} +\Delta _{n\textrm{d}} +\Delta _{n\textrm{c}} \, , \qquad
\Gamma _{n} =\Gamma _{nn} +\Gamma _{n\textrm{d}}
+\Gamma _{n\textrm{c}}\, ,
\end{equation}
where $\Delta _{nn}$ is the partial shift of the $n$-th state due to the
intra-level interaction within I-phonons
\begin{equation} \label{EQ:32}
\Delta _{nn} =\left(\frac{L}{2\pi } \right)^{2} \sum _{\lambda
,\pm }\mathcal P \int \int \rd^{2} \vec{q}(E_{n} -E_{n\vec{q}} \mp
\Omega _{\lambda \vec{q}} )^{-1}\left|F_{nn}^{(\lambda ,\vec{q})}
\right|^{2} \left(\nu _{\lambda \vec{q}} +\frac{1}{2} \pm
\frac{1}{2} \right),
\end{equation}
$\Gamma _{nn}$ is the decay rate of the $n$-th state due to the
intra-level interaction within I-phonons
\begin{equation} \label{EQ:33}
\Gamma _{nn} =\frac{L^{2}}{2\pi }\sum _{\lambda ,\pm }\int \int
\rd^{2} \vec{q} \left[\delta (E_{n} -E_{n\vec{q}} \mp \Omega
_{\lambda \vec{q}} )\right] \left|F_{nn}^{(\lambda ,\vec{q})}
\right|^{2} \left(\nu _{\lambda \vec{q}} +\frac{1}{2} \pm
\frac{1}{2} \right),
\end{equation}
$\Delta _{n{\left(\begin{smallmatrix}\textrm{c} \\ \textrm{d}
\end{smallmatrix}\right)}} $
and $\Gamma _{n{\left(\begin{smallmatrix}\textrm{c} \\ \textrm{d}
\end{smallmatrix}\right)}} $ are the partial shifts and decay rates
of the $n$-th state due to the interaction within I-phonons with all
(except the $n$-th) states of quasi-discrete (d) spectrum ($\Delta _{n\textrm{d}}
=\sum_{n_{1} \ne n}\Delta _{nn_{1} }$, \linebreak
 $\Gamma _{n\textrm{d}} =\sum_{n_{1} \ne n}\Gamma _{nn_{1} }  $)
and with the states of continuum (c) ($\Delta_{n\textrm{c}}=\sum_{k_{z} }
\Delta _{nk_{z} }$, $\Gamma_{n\textrm{c}}=\sum_{k_{z} }\Gamma _{nk_{z} }  $),
\begin{eqnarray}
\Delta _{n{\left(\begin{smallmatrix}\textrm{c} \\ \textrm{d}
\end{smallmatrix}\right)}} -\ri\, \Gamma _{n{\left(\begin{smallmatrix}\textrm{c} \\ \textrm{d} \end{smallmatrix}\right)}}
&=&\left(\frac{L}{2\pi } \right)^{2}
\sum _{\left(\begin{smallmatrix}k_{z} \\{n_{1} \ne n} \end{smallmatrix}\right)}
\sum _{\lambda ,\pm }\int \int \rd^{2} \vec{q}
F_{n\left(\begin{smallmatrix}n_{1} \\k_{z} \end{smallmatrix}\right)}^{*} (\lambda ,\vec{q})F_{n\left(\begin{smallmatrix}n_{1} \\k_{z} \end{smallmatrix}\right)} (\lambda ,\vec{q})\left(\nu _{\lambda \vec{q}} +\frac{1}{2} \pm \frac{1}{2} \right) \, \nonumber\\
\label{EQ:34}
&&\times \left[\, \mathcal P \, \left(E_{n}
-E_{n\left(\begin{smallmatrix}n_{1} \\k_{z} \end{smallmatrix}\right)} \mp \Omega _{\lambda \vec{q}} \right)^{-1}
-2\pi \ri\, \delta \left(E_{n} -E_{n\left(\begin{smallmatrix}n_{1} \\k_{z} \end{smallmatrix}\right)} \mp \Omega _{\lambda \vec{q}}
\right)\right].
\end{eqnarray}
Symbol $\mathcal P$ in formulas (\ref{EQ:32}), (\ref{EQ:34}) means that the
respective integrals are taken as Cauchy principal values.

Using the developed theory, we numerically calculated the energies renormalized due to
phonons ($\tilde{E}_{n} =E_{n} +\Delta _{n} $) and decay rates ($\Gamma _{n} $) of
electron states in quasi-discrete spectrum at the fixed physical and geometrical parameters of three-barrier RTS.

\section{Parameters of electron spectrum as functions of temperature and design of three-barrier RTS  (GaAs/Al$_{x}$Ga$_{1-x}$As)}
 The complete and partial shifts and decay rates of electron spectrum in three-barrier
 RTS were calculated for GaAs/Al$_{x}$Ga$_{1-x}$As nano-structure, being the active element
 of the experimentally investigated QCD \cite{2,3,17}. The physical parameters are presented in table~\ref{tbl1}.

\begin{table}[htb]
\caption{Physical parameters of nanostructures.} \label{tbl1}
\vspace{2ex}
\begin{center}
\renewcommand{\arraystretch}{0}
\begin{tabular}{|c||c|c|c|c|c|}
\hline
& $\varepsilon _\infty$& $\hbar \omega_\textrm{L}, $~{meV}& $\hbar \omega_\textrm{T}$, meV& $m_e/m$ &$U$, meV\strut\\
\hline
\rule{0pt}{2pt}&&&&&\\
\hline \raisebox{-0.2ex}[0pt]{GaAs}
      &  10.89&  36.25&  33.29&  0.067&  \strut\\
\hline \raisebox{-0.2ex}[0pt]{Al$_{0.15}$Ga$_{0.85}$As}
      & 10.48&  35.31&  33.19&  0.079&  120\strut\\
\hline \raisebox{-0.2ex}[0pt]{Al$_{0.45}$Ga$_{0.55}$As}
      & 9.66&  33.66&  32.77&  0.104&  320\strut\\
\hline
\end{tabular}
\renewcommand{\arraystretch}{1}
\end{center}
\end{table}

\begin{figure}[!t]
\centerline{
\includegraphics[width=0.8\textwidth]{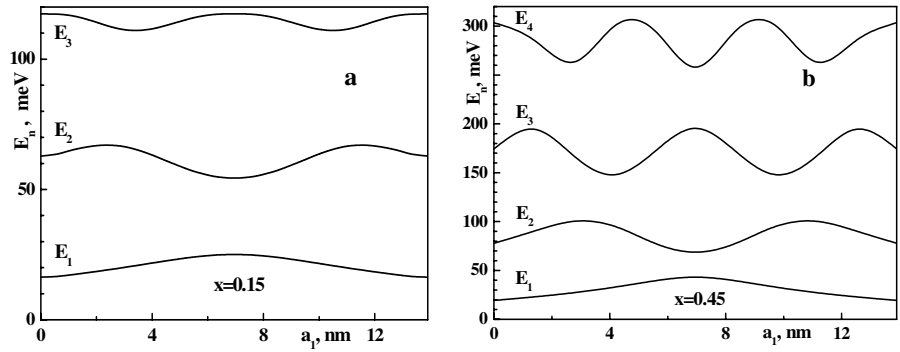}
}
\caption{Energy spectrum of electron ($E_{n}$) noninteracting
with phonons as a function of the inner barrier position ($a_{1}$) at
$x=0.15$ and $U=120$~meV (a) and $x=0.45$ and $U=320$~meV (b); $a=13.9$~nm,
$b=1.13$~nm.} \label{fig2}
\end{figure}

In figure~\ref{fig2}, the electron spectrum as the function of  the position of the inner barrier with respect to
the outer ones is presented at different Al concentrations: $x=0.15$ and low potential barrier
$U=120$~meV (a); $x=0.45$ and high potential barrier $U=320$~meV (b). The thicknesses of the inner
barriers ($b=1.13$~nm) and the sum of both well widths ($a=a_{1} +a_{2} =13.9$~nm) are the same for
the both structures. The figure proves that independently of Al concentration, the quasi-discrete
energy levels ($E_{n}$) qualitatively similarly depend on the position of the inner barrier
(fixed by the width of input well ($a_{1} $)): the energies $E_{n} $ are the symmetric functions
with respect to the average position of the inner barrier in the common well ($a_{1} =a_{2} =a/2$) with ${n}$ maxima.
Two nearest operating levels ($E_{1}$, $E_{2} $) have one maximum at $a_{1} =a/2$ and two maxima~--- at $a_{1} =a/4, \, 3a/4$,
respectively. The anti-crossing is observed at $a_{1} =a/2$, where the distance between $E_{2} $ and $E_{1} $
is minimal, due to the presence of two wells in a three-barrier RTS.

\begin{figure}[!b]
\centerline{
\includegraphics[width=0.8\textwidth]{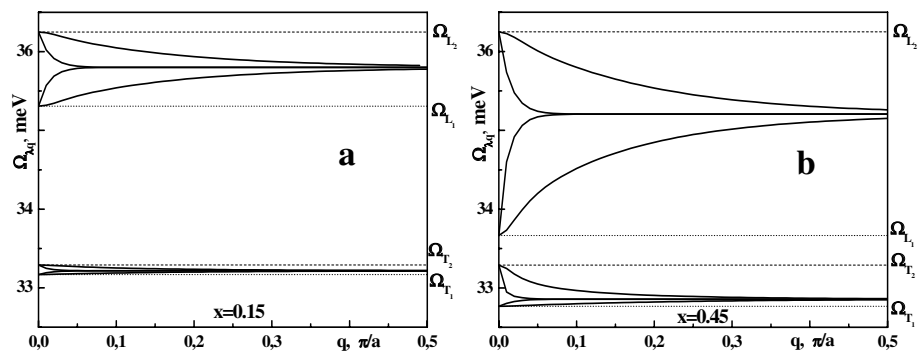}
}
\caption{Energy spectra ($\Omega _{\lambda \vec{q}} $) of
interface phonons uncoupling with electrons as a function of
quasi-momentum ($q$) at $x=0.15$ and $U=120$~meV (a) and $x=0.45$ and
$U=320$~meV (b); $a=13.9$~nm, $b=1.13$~nm; $\Omega _{\textrm{L}_{1}} $, $\Omega
_{\textrm{L}_{2}} $ are the energies of longitudinal phonons, $\Omega
_{\textrm{T}_{1}} $, $\Omega _{\textrm{T}_{2}} $ are the energies of transversal
phonons in the wells (1) and barriers (2).} \label{fig3}
\end{figure}

In figure~\ref{fig3}, the energy spectra of interface phonons ($\Omega
_{\lambda \vec{q}} $ )are presented for the both three-barrier RTS
(a), (b). The spectra contain two groups having four modes of
energies of a  weak dispersion. The high-energy group is placed
between the energies of longitudinal phonons ($\Omega _{\textrm{L}_{1} } $,
$\Omega _{\textrm{L}_{2} } $) and the low-energy group~--- between the
energies of transversal phonons ($\Omega _{\textrm{T}_{1} } $, $\Omega
_{\textrm{T}_{2} } $) of the respective layers. In each group, the pair of
modes with a higher energy has a negative dispersion while with lower
energy~--- positive dispersion. The position of the inner barrier
with respect to  the outer ones weakly effects
the magnitude of the dispersion: $\Omega _{\lambda \vec{q}} $
varies at 2--3~\% at big ${q}$ and the energies become almost
the same at a small ${q}$. An increase of  Al concentration in the
barriers does not qualitatively vary the dispersion of all phonon modes
 but increases its magnitude at a small quasi-momentum.

\begin{figure}[!b]
\centerline{
\includegraphics[width=0.75\textwidth]{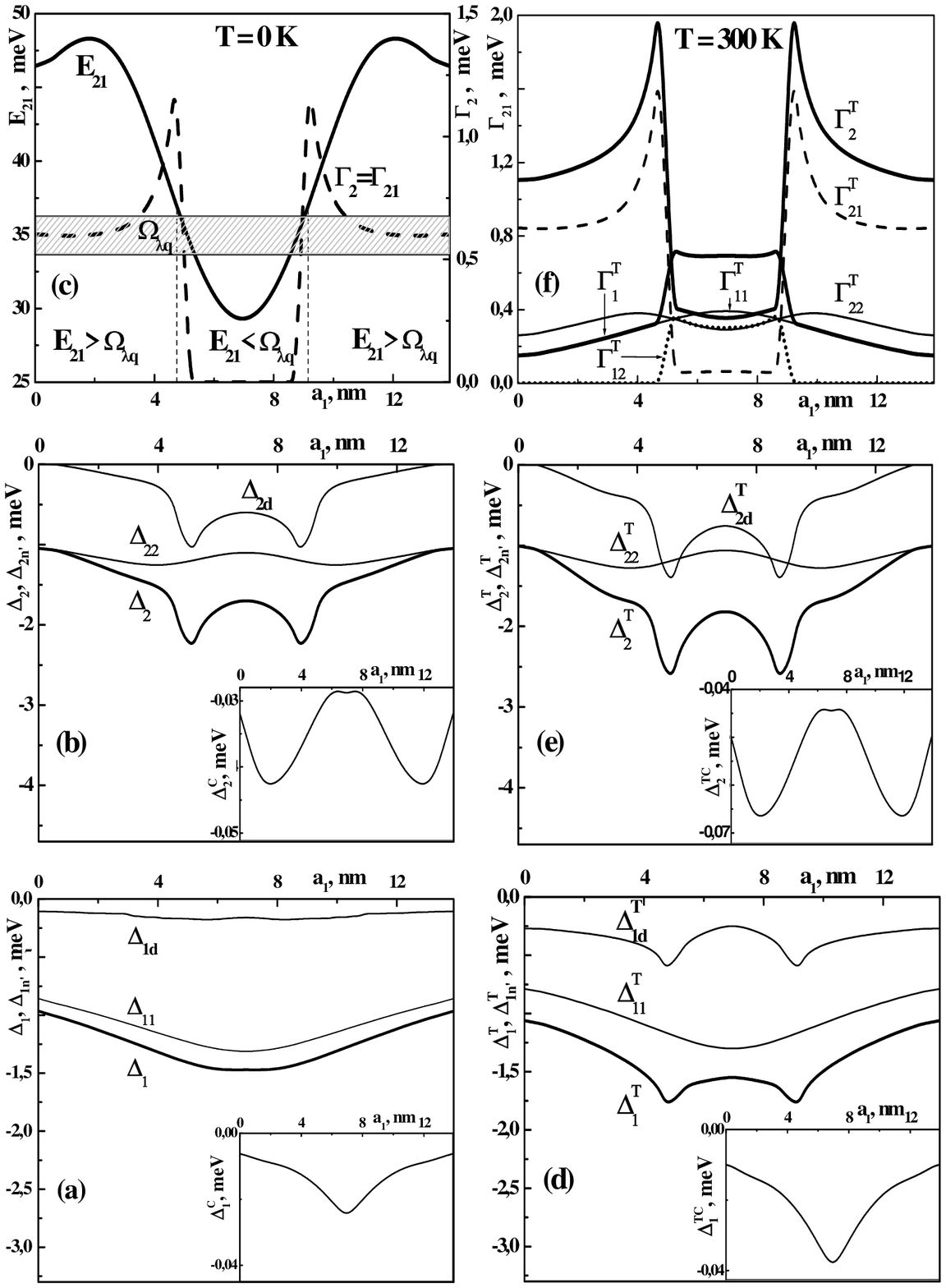}
}
\caption{Electron energy shifts and decay rates as a function of
the inner barrier position at $x=0.15$ and $T=0$~K (a), (b), (c) and $T=300$~K
(d), (e), (f); $a=13.9$~nm, $b=1.13$~nm. In figures~(d), (e), (f),
$\Delta_{1}^\textrm{T}$, $\Delta_{2}^\textrm{T}$ and $\Gamma_{1}^\textrm{T}$,
$\Gamma_{2}^\textrm{T}$ are the total shifts and decay rates of ($|1\rangle$)and ($|2\rangle$) states. $\Delta_{11}^\textrm{T}$,
$\Delta_{22}^\textrm{T}$ and $\Gamma_{11}^\textrm{T}$, $\Gamma_{22}^\textrm{T}$ are the
partial shifts and decay rates caused by intra-level interaction
due to I-phonons; $\Delta_{1\textrm{d}}^\textrm{T}$, $\Delta_{2\textrm{d}}^\textrm{T}$ and
$\Gamma_{1\textrm{d}}^\textrm{T}$, $\Gamma_{2\textrm{d}}^\textrm{T}$ are the partial shifts and
decay rates caused by inter-level interaction with the states of
discrete spectrum due to I-phonons; $\Delta_{1\textrm{c}}^\textrm{T}$,
$\Delta_{2\textrm{c}}^\textrm{T}$ and $\Gamma_{1\textrm{c}}^\textrm{T}$, $\Gamma_{2\textrm{c}}^\textrm{T}$ are the
partial shifts and decay rates caused by inter-level interaction
with the states of continuum due to I-phonons. In figures ~(a), (b), (c),
the same magnitudes calculated at $T=0$~K are presented without
index T.}\label{fig4}
\end{figure}

In order to study the effect of electron-I-phonon interaction on the magnitude of the
electromagnetic field energy absorbed by a nano-structure ( $\tilde{E}_{21} =\tilde{E}_{2} -\tilde{E}_{1} $)
arising due to the quantum transition from the first ($|1\rangle$) into the second ($|2\rangle$)
quasi-stationary state, we calculated the complete and partial shifts
($\Delta _{n}$, $\Delta _{n\textrm{d}}$, $\Delta _{n\textrm{c}}$) and decay rates
($\Gamma _{n},$ $\Gamma _{n\textrm{d}}$, $\Gamma _{n\textrm{c}} $) of two operating states ($n=1, 2$).
The results are presented in figure~\ref{fig4} at $T=0$ and 300~K for a nano-structure (a)
because Al concentration causes their small quantitative changes for a nano-structure (b).

Figure \ref{fig4} (a), (b) proves that at cryogenic temperatures (formally at $T=0$~K), the both
operating states ($|1\rangle$ and $|2\rangle$) shift into the region of smaller
energies ($\Delta _{1}$,  $\Delta _{2} <0$) independently of three-barrier RTS design
due to electron-I-phonon interaction. The magnitudes of complete shifts nonlinearly
depend on the position of the inner barrier ($a_{1} $) and are of the same order. The complete
shifts are mainly produced by intra-level interactions with partial shifts $\Delta _{11}$,  $\Delta _{22}$.
The shifts $\Delta _{1\textrm{d}}$, $\Delta _{2\textrm{d}} $ are produced by inter-level interaction due to
all states of a quasi-discrete spectrum and are smaller than $\Delta _{11}$, $\Delta _{22}$.
Only at some magnitudes of $a_{1} $, the partial shifts $\Delta _{22} $ and $\Delta _{2\textrm{d}} $
have correlating magnitudes. The partial shifts ($\Delta _{1\textrm{c}}$, $\Delta _{2\textrm{c}} $)
are caused by the interaction with the continuum states and are smaller or much smaller
than the others. The shifts are negative because at $T=0$~K only virtual phonons exist in the system.

The decay rate of electron spectrum ($\Gamma _{n} $) is regulated
by the energy conservation law which, as it is clear from MO (\ref{EQ:23}) at
$T=0$~K, is determined by $\delta $-function $\delta (E_{n}
-E_{n_{1}} -\Omega _{\lambda \vec{q}} -\hbar ^{2} q^{2} /2m)$.
Consequently, when for the $n$-th state the condition $n\leqslant n_{1} $
fulfills, the difference of energies becomes $E_{nn_{1} } =E_{n}
-E_{n_{1} } \leqslant 0$; thus, $\delta (-|E_{nn_{1} } |-\Omega
_{\lambda \vec{q}} -\hbar ^{2} q^{2} /2m)=0$ and, since, $\Gamma
_{n\leqslant n_{1} } =0$. As far as $E_{n} <E_{k_{z} } $, the same
reason brings to $\Gamma _{nk_{z} } =0$. Physically, it means that
the intra-level and inter-level interaction between electrons from
lower states and electrons from higher states of a discrete (d) and
a continuum (c) spectrum due to virtual I-phonons ($T=0$~K) occurs
without decay. When the electrons from higher ($n$) states interact
with virtual I-phonons through the lower ($n_{1} $) states, at
$n>n_{1} $, the difference of the energies $E_{nn_{1} } =E_{n}
-E_{n_{1} } >0$, thus at $E_{nn_{1} } <\Omega _{\lambda \vec{q}}
$, $\delta (E_{nn_{1} } -\Omega -\hbar ^{2} q^{2} /2m)=0$ and
$\Gamma _{nn_{1} } =0$ while at $E_{nn_{1} } >\Omega _{\lambda
\vec{q}} $, $\delta (E_{nn_{1} } -\Omega -\hbar ^{2} q^{2} /2m)\ne
0$ and $\Gamma _{nn_{1} } \ne 0$. In the last case, only the
inter-level interaction due to phonons causes the finite decay of
higher states at cryogenic temperatures ($T=0$~K).

Figure \ref{fig4} c presents the mentioned reasons determining the decay rates of electron
states at $T=0$~K. In the figure one can see the decay rate ($\Gamma _{2} =\Gamma _{21} $) of the
energy of quantum transition $E_{21} =E_{2} -E_{1} $ detected at the absorption of electromagnetic
wave, as a function of design of three-barrier RTS ($a_{1} $) and the stripe, where all modes of
I-phonon energies ($\Omega _{\lambda \vec{q}} $) are located. The figure proves that the decay rate of
the first state ($|1\rangle$) is absent ($\Gamma _{1} =0$) and, as far as $\Gamma _{22} =0$,
the decay rate of the second state ($|2\rangle$) is caused only by the inter-level interaction due
to phonons ($\Gamma _{2} =\Gamma _{21} $). Thus, $\Gamma _{2} =0$ in the range $4.75\leqslant a_{1} \leqslant 9.15$
where $E_{21} \leqslant \Omega _{\lambda q} $ and $\Gamma _{2} \ne 0$ in the ranges $0\leqslant a_{1} \leqslant 4.75$
and $9.15\leqslant a_{1} \leqslant 13.9$ where $E_{21} >\Omega _{\lambda q} $.

At the finite temperature, the average occupation number of phonon states is not equal to zero
($\nu _{\lambda \vec{q}} \ne 0$). Therefore, as it is clear from MO (\ref{EQ:23}), the spectral parameters
of both operating states are produced not only by virtual (as at $T=0$~K) but also by real I-phonons,
both in the processes of their creation, described by the first term of MO proportional to
($1+\nu _{\lambda \vec{q}} $) and in the processes of their annihilation, described by the
second term of MO proportional to $\nu _{\lambda \vec{q}} $.

In figure~\ref{fig4} (d), (e), (f), the complete and partial shifts ($\Delta _{n}^\textrm{T}$, $\Delta _{n\textrm{d}}^\textrm{T}$, $\Delta _{n\textrm{c}}^\textrm{T} $)
and decay rates ($\Gamma _{n}^\textrm{T}$, $\Gamma _{n\textrm{d}}^\textrm{T}$, $\Gamma _{n\textrm{c}}^\textrm{T}$) of the operating electron states ($n=1,2$)
are presented at $T=300$~K. It is clear that the temperature weakly changes the shapes and magnitudes of both
states shifts depending on the position of the inner barrier ($a_{1} $). The decay rates ($\Gamma _{1}^\textrm{T}$, $\Gamma _{2}^\textrm{T} $)
are produced by partial contributions of intra-level ($\Gamma _{11}^\textrm{T}$, $\Gamma _{22}^\textrm{T} $)
and inter-level ($\Gamma _{12}^\textrm{T}$, $\Gamma _{21}^\textrm{T} $) interactions with I-phonons in the processes
of their creation and annihilation.

\begin{wrapfigure}{i}{0.5\textwidth}
\centerline{
\includegraphics[width=0.48\textwidth]{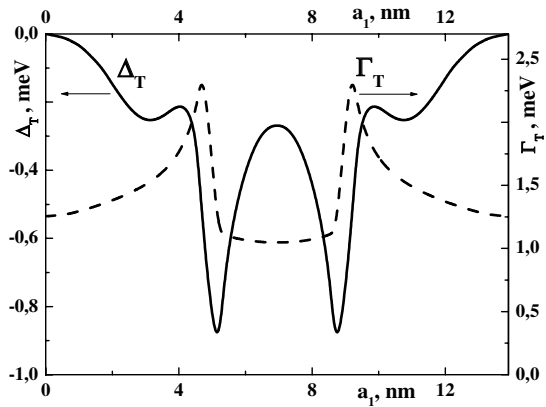}
}
\caption{A complete shift ($\Delta _\textrm{T}$) and the decay rate ($\Gamma
_\textrm{T}$) as a function of the inner barrier position ($a_1$) at $T=300$~K;
$a=13.9$~nm, $b=1.13$~nm.} \label{fig5}
\end{wrapfigure}

 In figure~\ref{fig5}, a complete shift ($\Delta _\textrm{T} =\Delta _{2}^\textrm{T} - \Delta _{1}^\textrm{T} $)
 and decay rate ($\Gamma _\textrm{T} =\Gamma _{1}^\textrm{T} + \Gamma _{2}^\textrm{T} $) of energy ($\tilde{E}_{21} =E_{21} +\Delta _\textrm{T} $)
 in the process of electromagnetic wave absorption as functions of the inner barrier position ($a_{1} $) is presented at $T=300$~K.
 The both functions are strongly nonlinear while the respective magnitudes are not big.

 We should note that sharp minima at the curves of all shifts and
 respective maxima at the curves of the decay rates (figures~\ref{fig4}, \ref{fig5}) are
 mainly caused by a bigger contribution of inter-level interaction
 between the second and third level in those two configurations of
 a three-barrier RTS where the anticrossing arises between them.

 According to physical considerations, the decay increases and the magnitude of the detected energy
 weakly decreases when the temperature increases, which qualitatively correlates with the experimental results \cite{17}.

 It is clear that one should consider in the model the confined polarization and acoustic phonons
 in order to quantitatively compare the theoretical and experimental data. This rather complicated
 and sophisticated work will be done in further investigations based on the approach proposed in this paper.

\section{Main results and conclusions}

 From the first principles (without any fitting parameters), we obtained the Hamiltonian of electron-I-phonon
 system in the representation of the second quantization over all variables for a three-barrier RTS. The renormalized
 electron spectrum was calculated at cryogenic and room temperatures using the thermo-dynamical Green's functions method.
 For GaAs/Al${}_{x}$Ga${}_{1-}$${}_{x}$As nano-structure, we studied in detail the effect of various mechanisms
 of electron-phonon interaction (intra-level and inter-level with quasi-discrete and continuum spectrum)
 at the formation of energy shifts and decay rates of electron states depending on the geometric design of three-barrier RTS.

 The energies of electron states, the energy of electromagnetic field absorbed by a three-barrier RTS which is
 an active element of QCD, at quantum transition between the first and the second states, the shifts of the energies
 and decay rates due to the electron-I-phonon interaction are not big over the magnitude, while nonlinear functions
 depend on the position of the inner barrier between the outer barriers.

 It is proven that independently of the geometric design of a three-barrier RTS, an increase of temperature causes
 bigger decay rates of both operating electron states and their shift into the low-energy region.
 The shift of the first level is smaller than that of the second one. Thus, being detected by QCD, the energy of
 electromagnetic field absorbed at a quantum transition decreases, which qualitatively correlates with the experiment.

%
%

\ukrainianpart

\title{Вплив інтерфейсних фононів на електронні робочі стани трибар'єрної резонансно-тунельної наноструктури як активної зони квантового каскадного детектора}
\author{М.В.Ткач, Ю.О.Сеті, Ю.Б. Гринишин, О.М.Войцехівська}
\address{Чернівецький національний університет ім. Ю.~Федьковича,\\ вул. Коцюбинського, 2, 58012 Чернівці, Україна}

\makeukrtitle

\begin{abstract}
\tolerance=3000%
З перших принципів у моделі ефективних мас та поляризаційного
континууму встановлено гамільтоніан системи електронів
взаємодіючих з інтерфейсними фононами у трибар'єрній
резонансно-тунельній структурі. Методом функцій Гріна розраховано
температурні зміщення й загасання найнижчих (робочих) електронних
станів у залежності від геометричної конфігурації наносистеми
GaAs/ Al$_{x}$Ga$_{1-x}$As як активної зони квантового каскадного
детектора. Встановлено, що незалежно від температури системи
енергія квантового переходу в процесах поглинання
електромагнітного поля є нелінійною слабозмінною функцією від
положення внутрішнього відносно зовнішніх бар'єра наносистеми.
\keywords резонансно-тунельна наноструктура, інтерфейсні фонони,
квантовий каскадний детектор

\end{abstract}

\end{document}